\documentclass[twocolumn,showpacs,preprintnumbers,prd,superscriptaddress,nofootinbib]{revtex4}

\usepackage{amsmath}
\usepackage{amsfonts}
\usepackage{amssymb}
\usepackage{graphicx}
\usepackage{hyperref}
\usepackage{booktabs}
\usepackage{cleveref}
\usepackage{color}

\begin{document}

\title{Probing observational bounds on scalar-tensor theories from standard sirens}

\author{Rocco D'Agostino}
\email{rdagostino@na.infn.it}
\affiliation{Dipartimento di Fisica, Universit\`a di Napoli “Federico II”, Via Cinthia, I-80126, Napoli, Italy.}
\affiliation{Istituto Nazionale di Fisica Nucleare (INFN), Sez. di Napoli, Via Cinthia 9, I-80126 Napoli, Italy.}

\author{Rafael C. Nunes}
\email{rafadcnunes@gmail.com}
\affiliation{Divis\~ao de Astrof\'isica, Instituto Nacional de Pesquisas Espaciais, Avenida dos Astronautas 1758, S\~ao Jos\'e dos Campos, 12227-010, SP, Brazil}

\begin{abstract}
 Standard sirens are the gravitational wave (GW) analog of the astronomical standard candles, and can provide powerful information about the dynamics of the Universe. In this work, we simulate a catalog with 1000 standard siren events from binary neutron star mergers, within the sensitivity predicted for the third generation of the ground GW detector called Einstein telescope. After correctly modifying the propagation of GWs as input to generate the catalog, we apply our mock data set on scalar-tensor theories where the speed of GW propagation is equal to the speed of light. As a first application, we find new observational bounds on the running of the Planck mass, when considering appropriate values within the stability condition of the theory, and we discuss some consequences on the amplitude of the running of the Planck mass. In the second part, we combine our simulated standard sirens catalog with other geometric cosmological tests (Supernovae Ia and cosmic chronometers measurements) to constrain the Hu-Sawicki $f(R)$ gravity model. We thus find new and non-null deviations from the standard $\Lambda$CDM model, showing that in the future the $f(R)$ gravity can be tested up to 95\% confidence level. The results obtained here show that the statistical accuracy achievable by future ground based GW observations, mainly with the ET detector (and planed detectors with a similar sensitivity), can provide strong observational bounds on modified gravity theories.
\end{abstract}


%---------------------------------------------------
\maketitle
%---------------------------------------------------
\section{Introduction}

After 20 years of research, the nature of the physical mechanism responsible for accelerating the Universe at late times is still an open question, and a large variety of cosmological models have been and are continually proposed in the literature to explain such observations (see \cite{DE_review,MG_review01, MG_review03} for review).
This is essentially due to the difficulty of discriminating among different scenarios that respond to the observations in the same way, leading to a theoretical degeneracy.

The observation of new astrophysical sources, through a direct manifestation of gravitational effects, can provide rich physical information about the nature of  gravity, which should play a key role to probe new (or rule out) additional gravitational degree(s) of freedom, or exotic forms of energy such as dark energy. The gravitational waves (GWs) issued by binary systems, such as binary black hole (BBH) and/or  binary neutron star (BNS) detected by LIGO/VIRGO, certainly open a new window to investigate fundamental physics in this direction. At present, catalogues of GWs from 10 BBH mergers and 1 BNS merger are available \cite{LIGO01}. The latter, the GW170817 event \cite{GW170817}, observed at $z \simeq 0.009$, has imposed strong constraints on modified gravity/dark energy models \cite{GW_MG01,GW_MG02,GW_MG03,GW_MG04,GW_MG041,GW_MG05}. Also, GW170817 was the first standard siren (the gravitational-wave analog of an astronomical standard candle) event to be catalogued, once its electromagnetic counterpart (GRB170817) was measured. These observations were also used to measure $H_0$ at the 12\% accuracy, assuming a fiducial $\Lambda$CDM  cosmology \cite{SS01}. We refer the reader to  \cite{H0_GW1,H0_GW2,H0_GW3} for proposals to use standard siren to measure $H_0$ with more accuracy.

Given the central importance of GW astronomy, beyond the present performance of the LIGO and Virgo interferometers, plans for construction of several GW observatory interferometers (on earth and in space) are currently in preparation, such like LIGO Voyager \cite{CE}, Cosmic Explore \cite{CE}, Einstein Telescope (ET) \cite{ET_design01,ET_design02}, LISA \cite{LISA}, DECIGO \cite{DECIGO}, TianQin \cite{TianQin}, among others, to observe GWs in the most diverse frequencies bands and different types of GW sources. In this paper, we are particularly interested to use the sensitivity predicted for the ET \cite{ET_design01,ET_design02}, which is a third-generation ground-based  detector and it is envisaged to be several times more sensitive in amplitude than the advanced ground-based detectors in operation, covering the frequency band range from 1 Hz to $10^4$ Hz. Also, the ET is expected to have signal-to-noise ratio (SNR) for BBH and BNS mergers several times larger than the current measures, as well as to observe hundreds, thousands of these events throughout the whole operational time. Several works have been done using the ET sensitivity to simulate GWs standard siren in order to investigate diverse aspects in cosmology \cite{ET01,ET02,ET03,ET04,ET05,ET06,ET07,ET08,ET09,ET10,ET11,ET12,ET13,ET14,ET15,ET16,ET17,ET18,ET19,ET20,ET21,ET22,ET23}.

In this work, we generate a simulated catalog with 1000 standard siren events from BNS mergers, from the ET power spectral density noise, in order to evaluate forecasting observational constraints on scalar-tensor theories where the speed of GW propagation is equal to speed of light. First, assuming a well-known parametric model for the running of the Planck mass, and assuming appropriate stability conditions on the theory, we find new observational bounds on the amplitude of the running of the Planck mass and we discuss its possible implications. In the second part, we apply our simulated standard siren data on $f(R)$ gravity given by the Hu-Sawicki model in order to find new observational limits on such model. In both analysis, we find that the parameters that characterize deviations from General Relativity (GR) may be non-null, within some statistical borders. 

The manuscript is organized as follows. In Section \ref{sec-model}, we set our theoretical framework to show how the GW propagation is modified from scalar-tensor theories. In Section \ref{Methodology}, we describe our methodology to generate standard siren mock catalogs. 
In Section \ref{alphaM0} and \ref{f(R)}, we present our main results. Finally, in Section \ref{Conclusions}, we outline our final considerations and future perspectives. 

Throughout the text, we use units such that $c=\hbar=1$, and $M_P=1/\sqrt{8\pi G}$ is the Planck mass. Moreover, we adopt the flat Friedmann-Lema\^itre-Robertson-Walker (FLRW) metric, $ds^2=-dt^2+a(t)^2\delta_{ij}dx^idx^j$, 
where $a$ is the scale factor, normalized to unity today. 
As usual notation, we denote by a subscript `0' physical quantities evaluated at the present time, and by the prime and dot symbols the derivatives with respect to the conformal time ($\tau$) and cosmic time ($t$), respectively, related by $dt=a d\tau$. Lastly, we express the Hubble constant $(H_0)$ results in units of km/s/Mpc.

\section{Modified Gravitational Wave Propagation in Scalar-Tensor Gravity}
\label{sec-model}

The Horndeski theories of gravity \cite{Horndeski, Deffayet} (see \cite{Horndeski_revisao01,Horndeski_revisao02} for a review) are the most general Lorentz invariant scalar-tensor theories with second-order equations of motion and where all matter is universally coupled to gravity. They include, as a sub set, the archetypal modifications of gravity such as metric and Palatini $f(R)$ gravity, Brans-Dicke theories, galileons, among others. The Horndeski action reads
\begin{equation}
\label{acao_geral}
 S =  \int d^4 x \sqrt{-g} \left[ \sum_{i=2}^{5} M_P^2 \mathcal{L}_i + \mathcal{L}_m \right],
\end{equation}
where $g$ is the determinant of the metric tensor, and
\begin{align}
\mathcal{L}_2 = &\ G_2(\phi, X), \\
\mathcal{L}_3 = &-G_3(\phi, X) \Box \phi, \\
\mathcal{L}_4 =& -G_4(\phi, X)R + G_{4,X} [( \Box \phi)^2 - \phi_{;\mu \nu} \phi^{;\mu \nu}], \\
\mathcal{L}_5 = &- G_5(\phi, X)G_{\mu \nu}\phi^{;\mu \nu} -
\dfrac{1}{6}G_{5,X}[(\Box \phi)^3  \\ 
&+ 2 \phi_{;\mu \nu}  \phi^{;\mu \sigma} \phi^{;\nu}_{;\sigma}
 - 3 \phi_{;\mu \nu} \phi^{;\mu \nu} \Box \phi].
\end{align}
Here, $G_i$ ($i$ runs over 2, 3, 4, 5) are functions of a scalar field $\phi$ and the kinetic term $X \equiv -1/2 \nabla^\nu \phi \nabla_\nu \phi $, and $G_{i,X} \equiv \partial G_i/\partial X$. For $G_2 = \Lambda$, $G_4 = M^2_P/2$ and $G_3 = G_5  = 0$, we recover GR with a cosmological constant. For a general discussion on the model varieties for different $G_i$ choices after GW170817, see \cite{Horndeski_revisao02}.

Recently, the GW170817 event together with the electromagnetic counterpart showed that the speed of GW, $c_T$, is very close to the speed of light for $z < 0.01$, i.e. $ |c_T/c - 1| < 10^{-15}$ \cite{GW170817}. In the context of Horndeski gravity, in order to explain these constraints, the only option is to consider $G_{4,X} \approx 0$ and $G_5 \approx constant$ in the action above. It is important to note that this restriction applies only to the local Universe ($\lesssim$ 40 Mpc). Thus, in principle, nothing prevents from considering the presence of these terms at redshifts larger than $z = 0.01$. In fact, only future measurements at high $z$ can confirm whether $c_T = c$. Here, we assume $c_T = c$, without loss of generality in the analysis we are going to develop. Under this condition, the GW propagation obeys the equation of motion \cite{Saltas04}
\begin{equation}
\label{h}
h''_{ij} + (2 + \alpha_M) \mathcal{H} h'_{ij} + k^2 h_{ij} = 0,
\end{equation}
where $h_{ij}$ is the metric tensor perturbation and $\mathcal{H}\equiv a'/a$ is the Hubble rate in conformal time. Moreover, $\alpha_M$ is the running of the Planck mass, which enters as a friction term responsible for modifying the amplitude of GWs acting as a damping term: 
\begin{equation}
\label{alphaM}
\alpha_M = \dfrac{1}{H M^2_{*}} \dfrac{dM^2_{*}}{dt},
\end{equation}
where $M_{*}$ is the effective Planck mass:
\begin{equation}
M^2_{*} = 2(G_4 - 2XG_{4X} + XG_{5 \phi} - \dot{\phi} H X G_{5X}),
\end{equation}
and $H\equiv \dot{a}/a$ is the Hubble parameter.
Following the methodology presented in \cite{Atsushi01} (see also \cite{Atsushi02, Atsushi03}), we can write a generalized GW amplitude propagation for scalar-tensor theories as
\begin{equation}
\label{hDT}
 h = e^{-\mathcal{D}} h_{GR},
\end{equation}
where
\begin{equation}\label{DD}
\mathcal{D} = \dfrac{1}{2} \int^{\tau} \alpha_M \mathcal{H} d\tau'.
\end{equation}
Note that due the condition $c_T = c$, that is, $G_{4,X} \approx 0$ and $G_5 \approx constant$, we do not have phase corrections in Eq.~(\ref{hDT}).
As the GW amplitude is inversely proportional to the distance, one can interpret the amplitude modification in Eq.~(\ref{hDT}) as a correction to the luminosity distance, defining an effective luminosity distance, or equivalently, an effective amplitude correction as \cite{Atsushi02, MG_amplitude01, MG_amplitude02}:
\begin{equation}
\label{dL_Gw}
d_L^{GW}(z) = d_L^{EM}(z) \exp \left[\dfrac{1}{2} \int_0^{z} \dfrac{dz'}{1 + z'} \alpha_M (z') \right],
\end{equation}
where $d_L^{EM}$ is the standard electromagnetic luminosity distance as a function of the redshift\footnote{The redshift is defined as $z=a^{-1}-1$.}:
\begin{equation}
\label{dL_em}
d_L^{EM}(z)  = (1 + z) \int_0^z \dfrac{dz'}{H(z')}.
\end{equation}
This generalization has been recently investigated in some contexts of modified gravity (see, e.g., \cite{Atsushi01,Atsushi02,Atsushi03,MG_amplitude01,MG_amplitude02,MG_dL01,MG_dL02,MG_dL03,MG_dL04,MG_dL05,MG_dL06,MG_dL07,MG_dL08}). 

\begin{figure*}
\begin{center}
\includegraphics[width=3.2in]{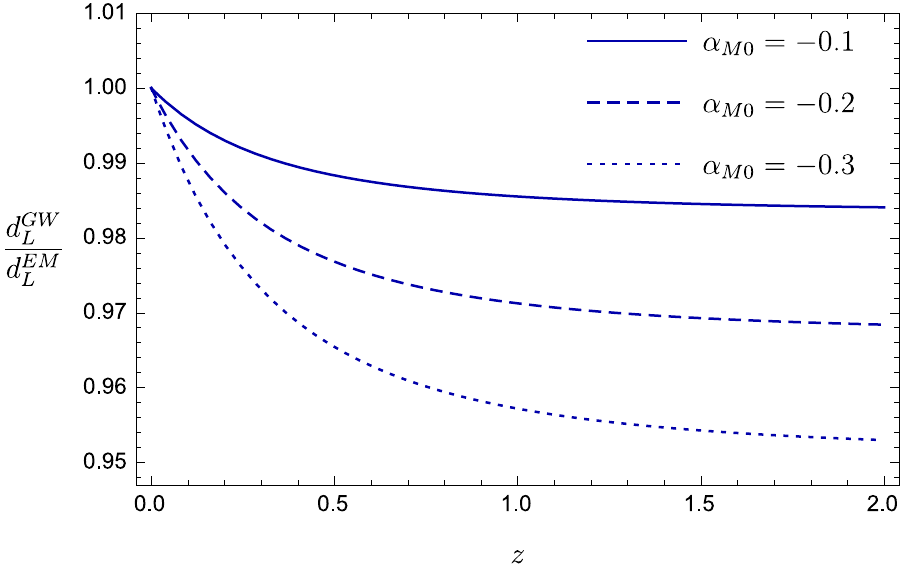} \,\,\,\,\,\,\,\,
\includegraphics[width=3.2in]{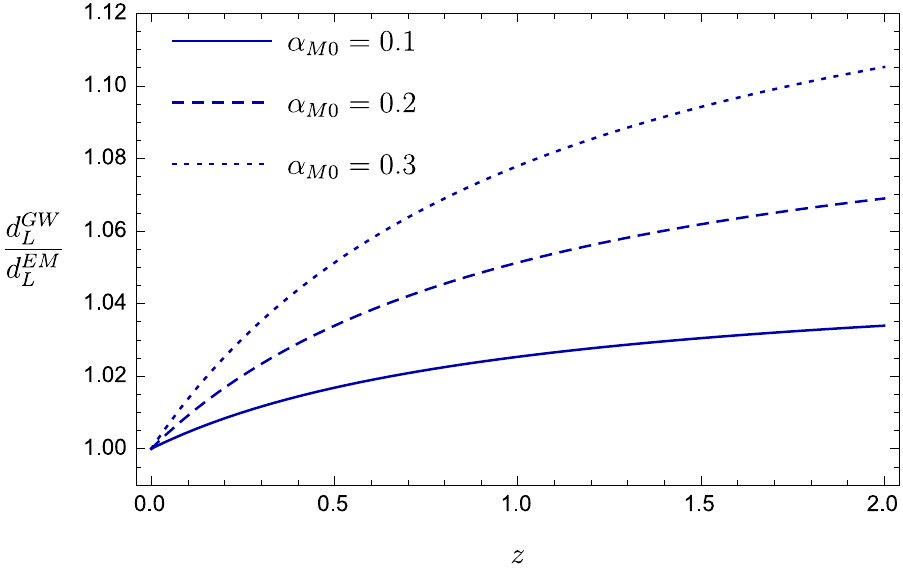}
\caption{Corrections on the effective luminosity distance (see Eq.~(\ref{dL_Gw})), as a function of the redshift, for different values of the running of the Planck mass today. The theoretical curves correspond to the case $\alpha_{M0} > 0$ and $n = 3$ (left panel), and $\alpha_{M0} > 0$ and $n=1$ (right panel).}
\label{fig:dL}
\end{center}
\end{figure*}

It is usual to choose phenomenologically motivated functional forms for the functions $\alpha_i$ (see, e.g., \cite{Bellini, alphai_01,alphai_02,alphai_03}). Typically, their evolution are tied to the scale factor $a(t)$ or to the dark energy density $\Omega_{de}(a)$ raised to some power $n$. On the other hand, an important point within Horndeski gravity are the stability conditions of the theory. Appropriate values of the free parameters functions must be considered in order to have a stable theory throughout the evolution of the Universe (see \cite{alphai_02} and reference therein). Following \cite{alphai_02}, we adopt the parametrization $\alpha_M =\alpha_{M0} a^n$, so that the stability conditions can be summarized as follows:
\begin{enumerate}
\item  $n > \dfrac{5}{2}$: stable for $\alpha_{M0} < 0$; 
\item  $0 < n < 1 + \dfrac{3\Omega_{m0}}{2}$: stable for $\alpha_{M0} > 0$.
\end{enumerate}
Here, $\Omega_{m0}$ is the present dimensionless matter density.
Under these considerations, we can note from Eq.~(\ref{dL_Gw}) that the changes in the GW amplitude propagation will be sensitive to the sign of $\alpha_{M0}$. Possible corrections with $\alpha_{M0} > 0$ or $\alpha_{M0}< 0$ will induce $d_L^{GW} > d_L^{EM}$ and $d_L^{GW} < d_L^{EM}$, respectively. We quantify these effects in Figure \ref{fig:dL}.  We note that variations on $\alpha_{M0} > 0 \, (< 0)$ can produce changes up to 10\% (5\%), respectively, on the effective GW amplitude, for reasonable values of the running of the Planck mass today. 

\section{Methodology and GW standard sirens data set}
\label{Methodology}

In order to move on, we need to define the GW signal $h_{GR}$. In modeling the gravitational waves form, given a GW strain signal $h(t) = A(t) \cos [\Phi(t)]$, we can obtain its Fourier transform $\tilde{h}(f)$ using the stationary phase approximation for the orbital phase of inspiraling binary system. For a coalescing binary system with component masses $m_1$ and $m_2$, we have
\begin{equation}
\label{waveform}
\tilde{h}(f) = Q \mathcal{A} f^{-7/6} e^{i\Phi(f)},
\end{equation}
where $\mathcal{A}$ is the GW inspiral amplitude computed perturbatively within the so-called post-Newtonian (PN) formalism up until 3 PN corrections, 
\begin{equation}
\label{A}
\mathcal{A} =  \sqrt{\dfrac{5}{96}} \dfrac{\mathcal{M}^{5/6}_c}{\pi^{2/3} d_L^{GW}} \left( \sum_{i=0}^6 A_i(\pi f)^{i/3} \right),
\end{equation}
where $d_L^{GW}$ is the modified luminosity distance as in Eq.~(\ref{dL_Gw}), and the coefficients $A_i$ are given in Appendix A.
The function $Q$ is expressed by
\begin{equation}
\label{Q}
Q^2 = F^2_{+}(1+cos^2(\iota))^2 + 2F^2_{\times}cos^2(\iota),
\end{equation}
where $\iota$ is the inclination angle of the binary orbital angular momentum with respect to the line of sight, and $F^2_{+}$, $F^2_{\times}$ are the two antenna pattern functions.
%typical of each specific GW detector.
In Eq.~(\ref{waveform}), the function $\Phi(f)$ is the inspiral phase of the binary system:
\begin{equation}
\label{phi}
\Phi(f) = 2 \pi f t_c - \phi_c - \dfrac{\pi}{4} + \dfrac{3}{128 \eta v^5} \left[ 1 + \sum_{i=2}^7 \alpha_i v^i  \right],
\end{equation}
where the coefficients $\alpha_i$ are the corrections up to the 3.5 PN corrections. In Appendix A, we also provide the expressions for these coefficients. 
In the above equation, we have defined $v \equiv (\pi M f)^{1/3}$, $M \equiv m_1 + m_2$, $\eta \equiv m_1 m_2/(m_1 + m_2)^2$, and $\mathcal{M}_c \equiv (1+z) M \eta^{3/5}$ to be the inspiral reduced frequency, total mass, symmetric mass ratio, and the redshifted chirp mass, respectively. The quantities $t_c$ and $\phi_c$ are the time and phase of coalescence, respectively.
\\

After having defined the modified GW signal for compact binaries, in what follows we  summarize the already know methodology used to estimate $d_L(z)$ measures from GW standard sirens. We refer to \cite{Schutz, Holz} for pioneer works in this regard. 

For a high enough signal-to-noise ratio (SNR) and a given waveform model, $h(f, \theta_i )$, with free parameters $\theta_i$,  we can use the Fisher matrix analysis to provide upper bounds for the free parameters of the models by means of the Cramer-Rao bound \cite{Fisher01, Fisher02}.  We refer the reader to \cite{Fisher03, Fisher04, Fisher05, Fisher06, Fisher07, Fisher08} for a discussion on the Fisher analysis to estimate parameters in binary systems for a given GW signal.  Once the waveform model is defined, the root-mean-squared error on any parameter is determined by
\begin{align}
\label{}
\Delta \theta^i = \sqrt{\Sigma^{ii}},
\end{align}
where $\Sigma^{ij}$ is the covariance matrix, i.e, the inverse of the Fisher matrix, $\Sigma^{ij} = \Gamma_{ij}^{-1}$. The Fisher matrix is given by
\begin{equation}
\label{}
\Gamma_{ij} = \left( \dfrac{\partial \tilde{h}}{\partial \theta^i} \mid  \dfrac{\partial \tilde{h}}{\partial \theta^j}  \right).
\end{equation}
The inner product between two waveform models is defined as
\begin{equation}
\label{}
(\tilde{h}_1 \mid \tilde{h}_2) \equiv 2 \int_{f_{low}}^{f_{upper}} \dfrac{\tilde{h}_1 \tilde{h}_2^{*} + \tilde{h}_1^{*} \tilde{h}_2}{S_n(f)} df,
\end{equation}
where the `star' stands for complex conjugation, and $S_n(f)$ is the detector spectral noise density. With this definition of the inner product, the SNR is defined as
\begin{equation}
{\rm SNR}^2 \equiv 4 Re \int_{f_{low}}^{f_{upper}} \, \dfrac{\vert h(f)\vert ^2}{S_n} df .
\label{eq:snr}
\end{equation}

In what follows, we consider the Einstein telescope (ET) detector power spectral density noise. The ET is a third-generation ground-based detector of GWs and it is envisaged to be ten times more sensitive in amplitude than the advanced ground-based detectors in operation nowadays, covering the frequency range $1-10^4$ Hz. Unlike the current detectors, from the ET conceptual design study, the expected rates of BNS detections per year are of the order of $10^3-10^7$~\cite{ET02}. However, we can expect only a small fraction ($\sim 10^{-3}$) of them accompanied with the observation of a short $\gamma$-ray burst. If we assume that the detection rate is in the middle range around $\mathcal{O}(10^5)$, we can expect to see $\mathcal{O}(10^2)$ events with short $\gamma$-ray bursts per year. 

Thus, let us consider in our simulations a mock GW standard sirens data set composed by 1000 BNS merger events.
Assuming that the errors on $d_L$ are uncorrelated with errors on the remaining GW parameters, we have
\begin{equation}
\sigma_{d_L}^2 = \left( \dfrac{\partial \tilde{h}(f)}{\partial d_L}, \dfrac{\partial \tilde{h}(f)}{\partial d_L} \right)^{-1}.
\end{equation}
Since $\tilde{h}(f) \propto {(d_L^{GW}})^{-1}$, then $\sigma_{d_L} =  d_L/ {\rm SNR}$. However, when we estimate the practical uncertainty of the measurements of $d_L$, we should take the orbital inclination into account. The maximal effect of the inclination on the SNR is a factor of 2 (between $\iota =0^{\circ}$ and $\iota = 90^{\circ}$). Therefore, we add this factor to the instrumental error for a conservative estimation. Thus, the estimate of the instrumental error is given by $\sigma_{d_L} =  2 d_L/ {\rm SNR}$. On the other hand, GWs are lensed in the same way as the electromagnetic waves, resulting into a weak lensing effect error, which we model as $\sigma^{\rm lens}_{d_L}= 0.05\ z\ d_L(z)$ \cite{ET02, ET_lens}. In our study, we do not consider possible errors induced from the peculiar velocity due to the clustering of galaxies. Since we are interested in simulating events at high $z$ mainly, we can neglect such contributions, which are significant only for $z \ll 1$.
In fact, at high $z$, the dominant source of uncertainty is the one due to weak lensing. Therefore, the total uncertainty $\sigma_{d_L}$ on the luminosity distance measurements associated to each event is obtained by combining the instrumental and weak lensing uncertainties as
\begin{align}
\sigma_{d_L} &=  \sqrt{ \left ( \sigma^{\rm ins}_{d_L} \right )^2 + \left ( \sigma^{\rm lens}_{d_L} \right )^2} \nonumber \\
&= \sqrt{ \left ( \dfrac{2d_L(z)}{\rm SNR} \right )^2 + (0.05 z d_L(z))^2}\,.
\label{eq:sigmatot}
\end{align}
The redshift distribution of the BNS sources is taken to be of the form
\begin{equation}
P(z) \propto \dfrac{4 \pi d^2_C(z) r(z)}{H(z)(1+z)},
\end{equation}
where $d_C(z)$ is the comoving distance, and $r(z)$ describes the time-evolution of the burst rate:
\begin{equation}
r(z)=\left\{
\begin{aligned}
& 1 + 2z, \hspace{0.2cm} z \leq 1 \\
& (15 - 3z)/4, \hspace{0.2cm} 1 < z < 5 \\
& 0,\hspace{0.2cm} z \geq 5 
\end{aligned}
\right.
\end{equation}
The distribution of the neutron star masses is chosen to be randomly sampled from uniform distributions within $[1 - 2]~{\rm M}_{\odot}$, also under the condition $m_1 \gtrsim m_2$ and $\eta < 0.25$. In this case, in the mock data generation we take $\chi_1 = \chi_2 = 0$, where $\chi_1, \chi_2$ are the associated spin magnitudes on each mass component. Then, we simulate BNS mergers up to $z =2$, which represents the maximum distance at which these events can be observed from the power spectral density noise from the ET \cite{ET02}. Also, we checked that, beyond $z = 2$, the SNR presents low values, which also can limit the use of the Fisher information for mock data. Now, in order to realistically generate a mock catalog using modified gravity, we shall consider non-zero values for the parameters $\alpha_{M0}$ and $n$, which are compatible with the current cosmological observation as well as with the stability criteria of the theory. Lastly, when generating our mock GW standard sirens data set, we only consider BNS mergers with ${\rm SNR} > 8$.

In order to estimate the observational constraints on the free parameters of the models, we apply Markov Chain Monte Carlo (MCMC) method through the Metropolis-Hastings algorithm \cite{Metropolis-Hastings}, where the likelihood function for the GW standard sirens mock data set is built in the form
\begin{equation}
\mathcal{L}_\text{GW} \propto \exp \left[ -\dfrac{1}{2} \sum_{i=1}^{1000} \left( \dfrac{d_L^{obs}(z_i) - d_L^{th}(z_i)}{\sigma_{d_L,i}}\right) \right].
\end{equation}
Here, $d_L^{obs}(z_i)$ are the 1000 simulated BNS merger events with their associated uncertainties $\sigma_{d_L,i}$, while  $d_L^{th}(z_i)$ is the theoretical prediction on each $i$th event.

\section{Constraints on the running of the Planck mass}
\label{alphaM0}

\begin{table*}
\begin{center}
\setlength{\tabcolsep}{1em}
\renewcommand{\arraystretch}{2}
\begin{tabular}{c c c c c}
\hline
\hline
Stability Conditions & $H_0$ & $\Omega_{m0}$ & $\alpha_{M0}$ \\
\hline
$\alpha_{M0}<0$ & $67.466^{+0.036(0.143)}_{-0.067(0.179)}  $ &  $0.328^{+0.015(0.028)}_{-0.014(0.028)} $ & $-0.100^{+0.051(0.092)}_{-0.043(0.085)} $  \\
$\alpha_{M0}>0$ & $67.390_{-0.050(0.095)}^{+0.047(0.098)}$ & $0.297^{+0.029(0.083)}_{-0.044(0.072)} $  &$0.199^{+0.069(0.178)}_{-0.097(0.167)}   $   \\
\hline
\hline
\end{tabular}
\caption{Summary of the MCMC results for the cases $\alpha_{M0} > 0$ and $\alpha_{M0}< 0$. The upper and lower values next to the mean value of each parameter denote the 68\% and 95\% CL errors.}
\label{tab:results}
\end{center}
\end{table*}

\begin{figure*}
\begin{center}
\includegraphics[width=3.2in]{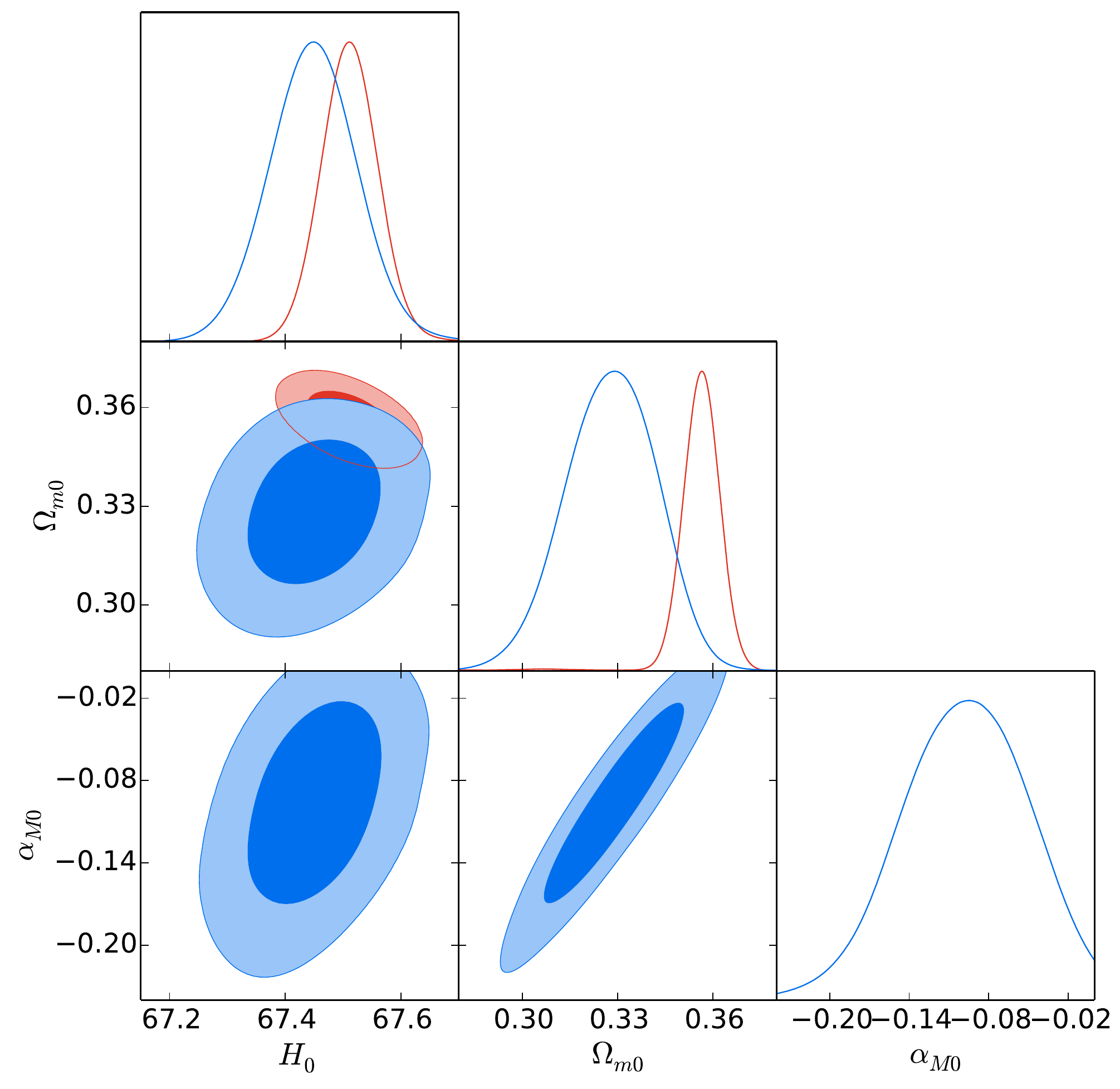} \,\,\,\,\,\,\,\,
\includegraphics[width=3.2in]{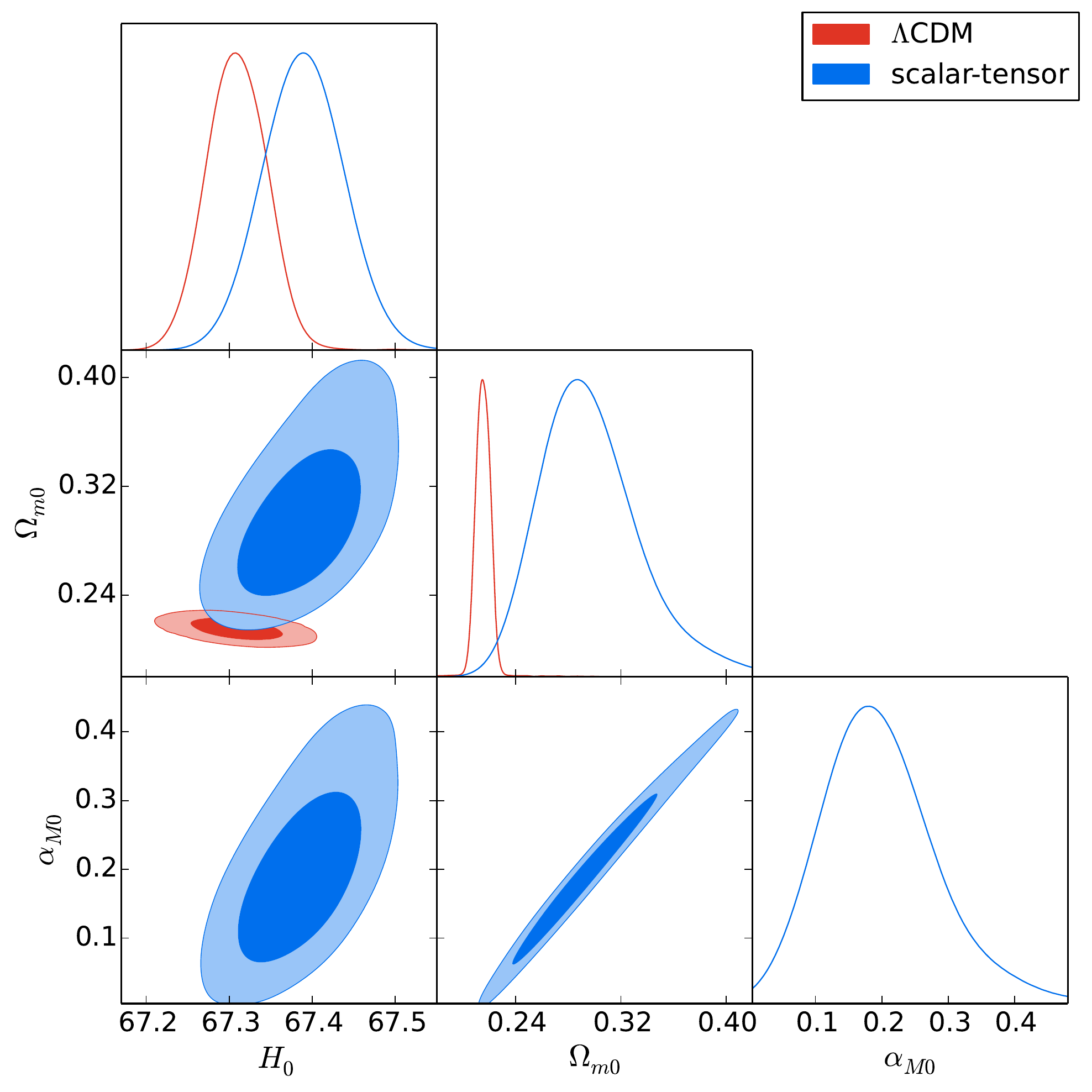}
\caption{Parametric space at 68\% and 95\% CL and one-dimensional marginalized distribution of $\Omega_{m0}$, $H_0$ and $\alpha_{M0}$ for scalar-tensor theories, resulting from the mock GW data generated under the stability condition $\alpha_{M0} < 0$ (left panel) and $\alpha_{M0} > 0$ (right panel). The predictions of the $\Lambda$CDM model within each forecast analysis are shown for comparison.}
\label{fig:contours_alphaM0}
\end{center}
\end{figure*}

In this section, we present and discuss our results regarding the future observational constraints that GW standard sirens can impose on a possible time variation of the Planck mass within of the ET sensitivity. The running of the Planck mass is an important physical quantity, which essentially is present in any and all modified gravity models. To generate a simulated $d_L(z_i)$ catalog using modified gravity, we assume realistic values for the pair ($\alpha_{M0}$, $n$), on each triplet ($z_i, d_L(z_i), \sigma_{d_L(z_i)}$) evaluated at each point $i$, as follows.

We first note that the parameter $n$ is statistically degenerate. This fact is already well known and expected to happen. In the literature, it is usual to assume $n = 1$, but here we follow the stability conditions discussed in Section \ref{sec-model} and weakly generate random values for $n$, within the range of stability of the theory: $i$) for the case $\alpha_{M0} > 0$, we randomly sampled from uniform distributions: $\alpha_{M0} \in  [0, 0.5]$ and $n \in  [0, 1.40]$; $ii$) for the case $\alpha_{M0} < 0$, we randomly sampled $\alpha_{M0} \in  [-0.5, 0]$ and $n \in  [2.5, 3.5]$. We found that different prior ranges on $n$ change the simulated catalogs very weakly. Only very different prior ranges on $\alpha_{M0}$ can significantly change the pair ($d_L(z_i), \sigma_{d_L(z_i)}$). The range assumed on $\alpha_{M0}$ is fully compatible with current constraints \cite{Horndeski_constraints_01,Horndeski_constraints_02,Horndeski_constraints_03,Horndeski_constraints_04,Horndeski_constraints_05,Horndeski_constraints_06,Horndeski_constraints_07,Horndeski_constraints_08,Horndeski_constraints_09,Horndeski_constraints_10,Horndeski_constraints_11}. We used as input values $H_0 = 67.4$ km/s/Mpc and $\Omega_{m0} = 0.31$ for the Hubble constant and matter density parameter, respectively, in agreement with the most recent Planck CMB data \cite{Planck2018}. Hence, these values are reasonable for our purpose to generate GW standard sirens mock data.

In the realization of the MCMC analysis, the sampling has been done assuming the following uniform priors for the cosmological parameters: $H_0 \in [55, 90]$, $\Omega_{m0} \in [0, 1]$, and $\alpha_{M0} \in [-1, 0]$, $\alpha_{M0} \in [0, 1]$ for each case. Due to the large statistical degeneracy on $n$, as commented above, we fixed $n= 3$ and $n=1$ for the cases of $\alpha_{M0} > 0$ and $\alpha_{M0}< 0$, respectively.
Table \ref{tab:results} summarizes the constraints at the 68\% and 95\% confidence levels (CL). In Figure \ref{fig:contours_alphaM0}, we show the parametric space and the one-dimensional marginalized distribution for the parameters $\Omega_{m0}$, $H_0$ and $\alpha_{M0}$, in both $\alpha_{M0} > 0$ and $\alpha_{M0}< 0$ cases. We note from both analyses that the parameter $\alpha_{M0}$ is non-null at the 68\% CL. Assuming the stability condition where $\alpha_{M0}$ is negative, we find the new lower limit $\alpha_{M0} > -0.2$ at the 95\% CL. On the other hand, under the condition that the running of the Planck mass is positive defined, we find that $\alpha_{M0}$ can be non-null up to 95\% CL, more specifically $0.03 \lesssim \alpha_{M0} \lesssim 0.38$. 
For a qualitative comparison, we show also the constraints from the $\Lambda$CDM model in both cases. It is important to note that the $\Lambda$CDM model is a particular case of our simulated data, where both mock catalogs are mainly controlled by possible corrections on the amplitude of the running of the Planck mass as explained above. Therefore, extended estimates on the pair ($d_L(z_i), \sigma_{d_L(z_i)}$) beyond the $\Lambda$CDM prediction will be dependent on the $\alpha_{M0}$ correction, where we have considered two different scenarios  $\alpha_{M0} > 0$ and $\alpha_{M0} < 0$. This should induce a minimal bias when analyzing the parameters within the $\Lambda$CDM scenario due to the $\alpha_{M0}$ prior range in the mock data generation. In the case $\alpha_{M0} < 0$, our minimal baseline, i.e. $\Omega_{m0}$ and $H_0$, is completely compatible for the two scenarios. On the other hand, when analyzing $\alpha_{M0} > 0$, we can note a minimal bias manifested on $\Omega_{m0}$, resulting in a minimum tension at 1$\sigma$ CL on this parameter, but still making the model compatible beyond that statistical significance.

It is interesting to compare our results with others already obtained in the literature. For instance, in \cite{Horndeski_constraints_11}, using measurements of the growth rate of structures from DESI survey, it is observed that the amplitude of the running of the Planck mass (quantity physically analogous to our $\alpha_{M0} > 0$) can be detected up to 99\% CL. In \cite{Horndeski_constraints_08}, a 95\% CL upper limit of 0.015 is found from CMB data. In \cite{MG_dL01}, analyzing the standard siren GW17081 event, the authors found the amplitude of the running of the Planck mass to be $\in$ [-80, 28] at the 95\% CL. In \cite{Atsushi02}, the amplitude damping $\alpha_{M0} < 0$ appear to be preferentially at low $z$ from GWs observations. Other analyzes can be found in \cite{Horndeski_constraints_01, Horndeski_constraints_02, Horndeski_constraints_03, Horndeski_constraints_04, Horndeski_constraints_05, Horndeski_constraints_06, Horndeski_constraints_07, Horndeski_constraints_08, Horndeski_constraints_09,Horndeski_constraints_10,Horndeski_constraints_11}. We note that the new borders on the amplitude of the running of the Planck mass derived in this work may have also an impact on the modified propagation primordial gravitational waves spectrum  \cite{Horndeski_constraints_12}.
\\

\begin{figure*}
\begin{center}
\includegraphics[width=3.2in]{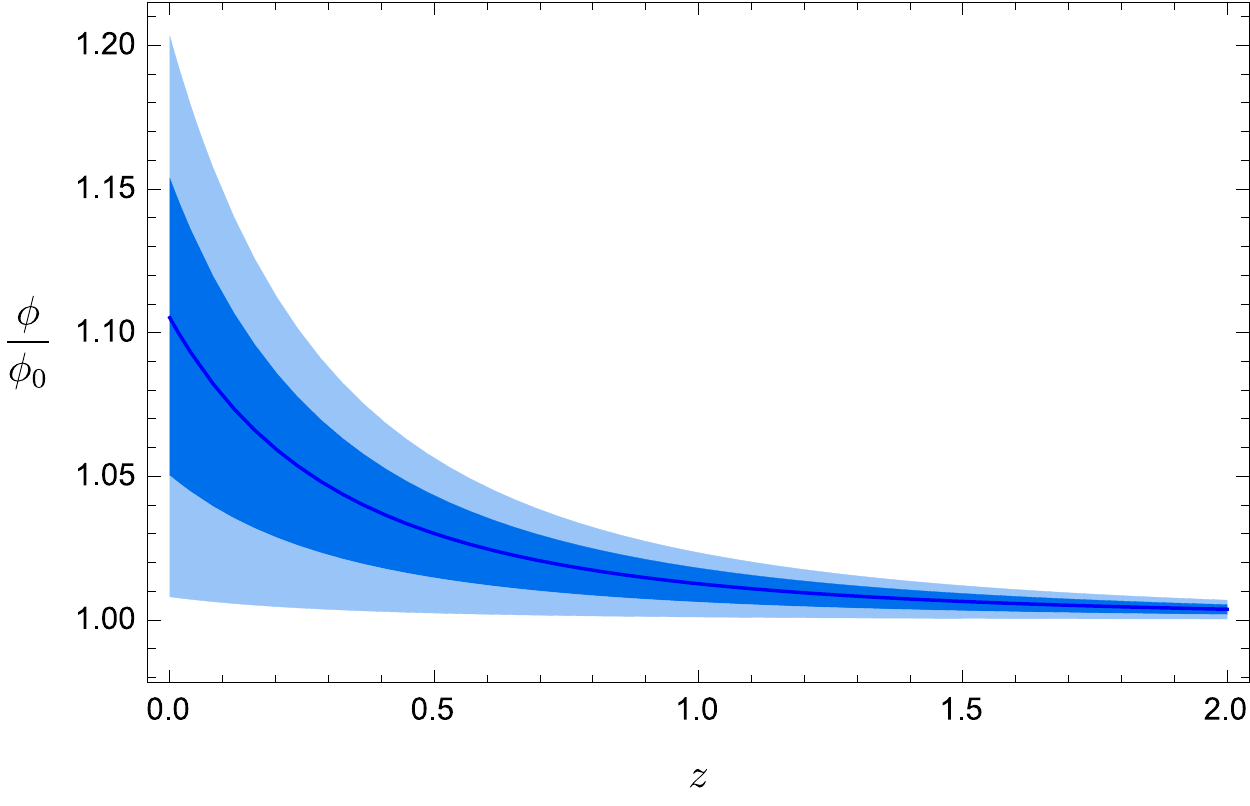} \,\,\,\,\,\,\,\,
\includegraphics[width=3.2in]{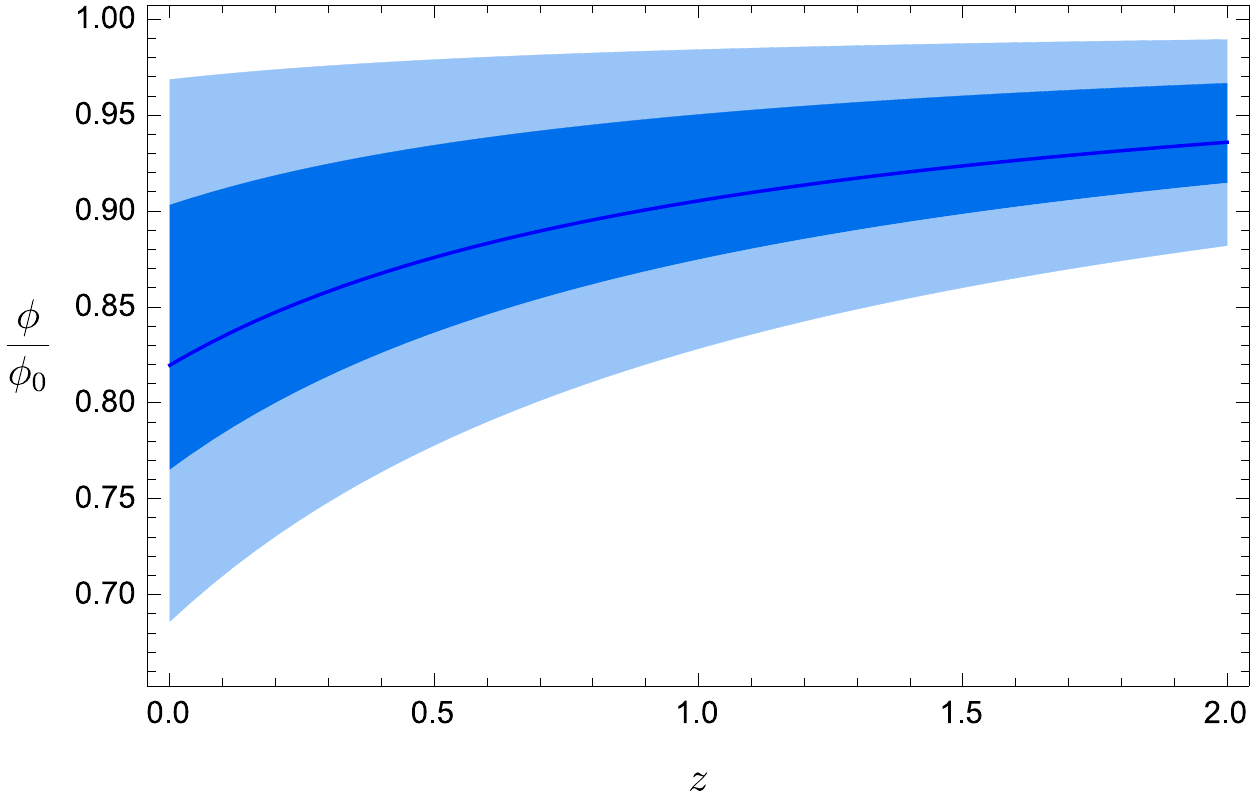}
\caption{Statistical reconstruction of the $\phi/\phi_0$ in Planck mass units as a function of $z$ at the 68\% CL and 95\% CL, under the stability condition $\alpha_{M0} < 0$ (left panel), and $\alpha_{M0} > 0$ (right panel).}
\label{fig:phi}
\end{center}
\end{figure*}

Now, we shall briefly discuss the consequences of our results. Based on the arguments developed in Section \ref{sec-model}, we can write the running of the Planck mass as
\begin{equation}
\label{running planck mass}
\alpha_M = \dfrac{\dot{G_4}}{H G_4}.
\end{equation}
One of the surviving classes of models under the condition $c_T = c$ are the non-minimal theories in which the scalar field $\phi$ is coupled with the curvature scalar $R$ in the form $G_4(\phi) R$. This class includes the metric $f(R)$ gravity and the Brans-Dicke theory \cite{Brans61}. The original Brans-Dicke theory, for instance, is obtained by setting $G_4 = \phi$. By substituting this in  Eq.~(\ref{running planck mass}), it is possible to obtain $\phi$ as a function of the cosmic time. In Figure \ref{fig:phi}, we show a reconstruction for the evolution of the field $\phi/\phi_0$ in Planck mass units, where $\phi_0$ is $\phi(z = 0)$.
Also, it is important to remember that in such a theory the gravitational constant is not presumed to be constant, but $G_N(\phi) \varpropto 1/\phi$. This fact is physically encoded in $\alpha_M(a)$, which measures the gravity strength. We can note that at late times, the gravity strength, $G_N$, is greater (smaller) than predicted by GR when $\alpha_{M0} > 0 \, (< 0)$, up to 20\% (30\%) at $z = 0$. On the other hand, when $z \gg 0$, GR is recovered and we do not expect to have significant variations at early times.

\section{Constraints on parametric $f(R)$ gravity}
\label{f(R)}

In this section, we briefly review $f(R)$ cosmology and show new observational constraints on the Hu-Sawicki (HS) model from our standard siren mock data set. We refer to \cite{fR1,fR2,Capozziello11,Nojiri17} for reviews on $f(R)$ gravity.

The $f(R)$ gravitational theories consist of extending the Einstein-Hilbert action in the form
\begin{equation}
S = \int d^4 x \sqrt{-g}\,\, \dfrac{M_P^2}{2}f(R) + S_m ,
\label{action0}
\end{equation}
where $f(R)$ is a function of the Ricci scalar, and $S_m$ is the action for matter fields. For $f(R) = R$, the GR case is recovered.

Let us consider a spatially flat FLRW Universe dominated by pressureless matter (baryonic plus dark matter) and radiation with energy densities $\rho_m$, $\rho_r$ and pressures $P_m$, $P_r$, respectively. The modified Friedmann equations in the metric formalism are given by \cite{fR1,fR2}
\begin{eqnarray}
3FH^2=8\pi G  \left(\rho_m+\rho_r\right) +\dfrac{1}{2} \left( FR - f \right)-3H\dot{F}\,, \label{FR1a} \\
-2F\dot{H} = 8\pi G  \left( \rho_m + P_m +\rho_r + P_r \right)+\ddot{F}-H\dot{F}\
\label{FR2a}
\end{eqnarray}
Moreover, one obtains the following useful relation:
\begin{equation}
R=6\left(2H^2+\dot{H}\right).
\end{equation}
It can be shown that, through the transformation of the scalar degree of freedom, $\phi = M_{P} df(R)/dR$, the metric $f(R)$ gravity is equivalent to the Brans-Dicke theory (with $w_{BD} = 0$). Thus, when considering a parametric function $f(R)$, given our constraints on $\phi$, some bounds can also be found on $f(R)$ gravity. For example, adopting the formalism presented in \cite{Basilakos, Rafael} (and reference therein), one can write
\begin{equation}
\label{fR_models}
f(R) = R - 2 \Lambda y(R, b),
\end{equation}
where the function $y(R, b)$ quantifies the deviation from Einstein gravity, i.e.  the effect of the $f(R)$ modification, through the distortion parameter $b$. Then, the scalar field can be expressed as
\begin{equation}
\phi \simeq M_{P} \left[1 - 2 \Lambda \dfrac{\partial y(R, b)}{\partial R} \right].
%\phi \simeq M_{pl} \left[ 1 - \dfrac{2 b}{b^2 + 2b/\Lambda + R^2/\Lambda}  \right]
\end{equation}
Interpreting $\phi_0 = M_{P}$ and given a function $y(R, b)$, we can use our limits on $\phi/\phi_0$ and place observational bounds on $b$.  We note that the stability conditions assumed in Section~\ref{sec-model} and used in the development of this section are completely in agreement with $f(R)$ gravity \cite{alphai_02}. Clearly, more direct observational boundaries on $f(R)$ gravity can be obtained by modifying appropriately Eq.~(\ref{dL_Gw}) to include a function $f(R)$. We are motivated to present a more detailed study in this sense in a future communication.

Without loss of generality, we can use our simulated standard siren catalog from BNS mergers, within a parametric limit and certain stability conditions, to model the $f(R)$ gravity dynamic. Thus, in what follows, we consider a parametric $f(R)$ gravity scenario and investigate the observational bounds that our standard siren mock data set can impose on the free parameters of the theory.

Let us consider viable models that have up to two parameters, where $f(R)$ function is given by Eq.~(\ref{fR_models}). This methodology has been used earlier to investigate the observational constraints on $f(R)$ gravity in \cite{Basilakos, Rafael}. In this respect, one of the most well-known models in the modified gravity theory literature is the Hu-Sawicki (HS) model \cite{Hu:2007nk}, which satisfies all the dynamics conditions required for a given $f(R)$ function. The function $y(R, b)$ for the HS model is given by
\begin{equation}
\label{y_HS}
y(R, b) = 1- \dfrac{1}{1+ \Bigl(\dfrac{R}{\Lambda b} \Bigr)^n},
\end{equation}
where $n$ is an intrinsic parameter of the model. In what follows, we assume $n = 1$ and refer to \cite{Basilakos, Rafael} for details.

\begin{figure}
\begin{center}
\includegraphics[width=3.2in]{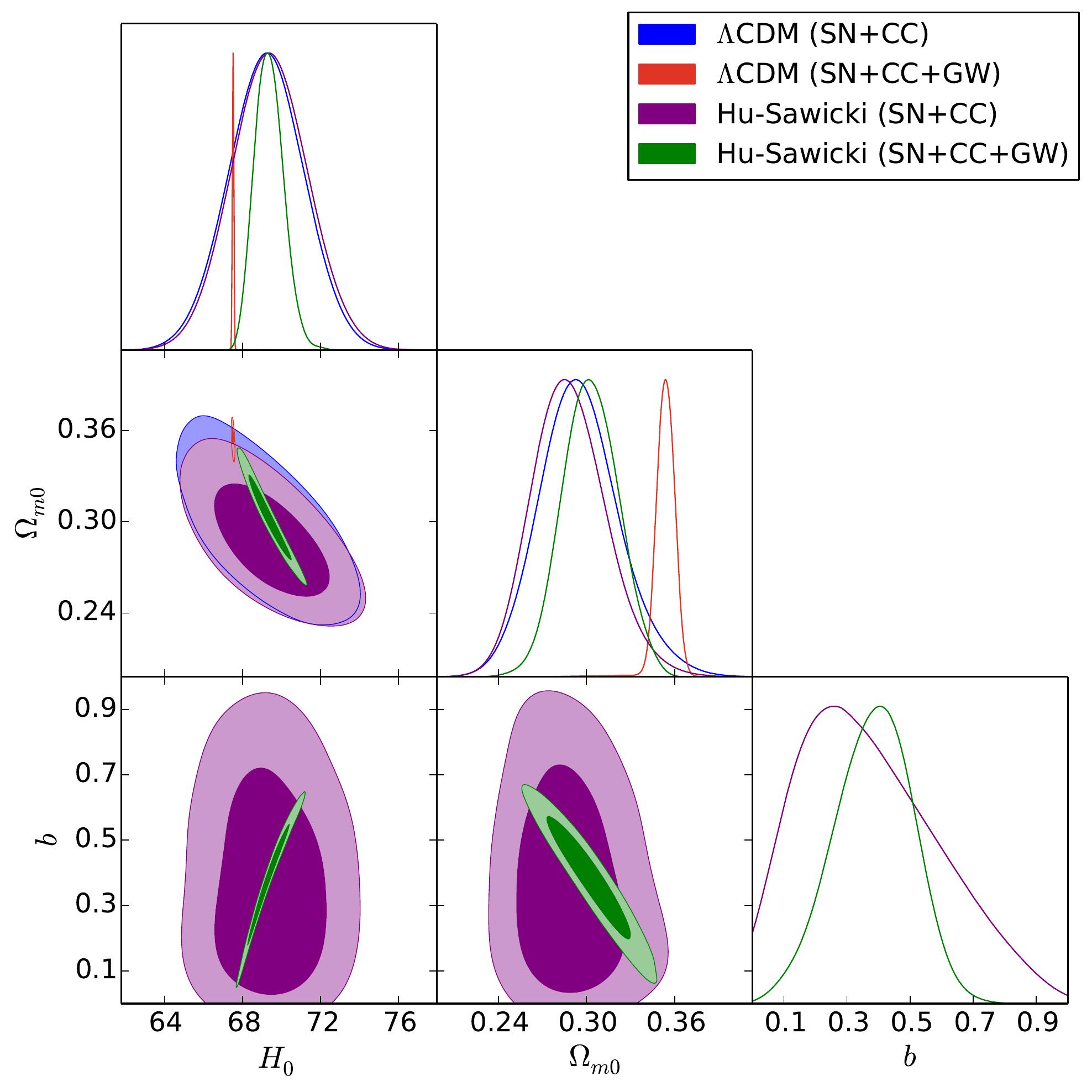}
\caption{Parametric space at 68\% and 95\% CL and the one-dimensional marginalized distribution for $\Omega_{m0}$, $H_0$ and $b$, for the Hu-Sawicki model from the joint analysis SNe + CC + GW and SNe + CC. The predictions of the $\Lambda$CDM model are shown for comparison.}
%($\alpha_{M0}<0$) data.}
\label{fig:contours_Hu-Sawicki}
\end{center}
\end{figure}

As a direct application of the standard siren events from BNS mergers, within the sensitivity predicted for the ET, we proceed to constrain the HS model using our mock data set generated by the condition $\alpha_{M0} < 0$. Together with the standard siren data, in the present analysis we also employed the type Ia Supernova (SN) Pantheon data \cite{Scolnic18} and the cosmic chronometers (CC) measurements \cite{Moresco12} in order to obtain tighter constraints on the free parameters of the HS scenario (see Appendix B for the details on the SNe and CC data.)
In the MCMC analysis, we assumed uniform priors for the cosmological parameters: $H_0 \in [55, 90]$, $\Omega_{m0} \in [0, 1]$, and $b \in [0, 1]$.
In Figure \ref{fig:contours_Hu-Sawicki}, we show the parametric space and the one-dimensional posterior distributions for the HS baseline model from SN + CC + GW and SN + CC analysis. Thus, we can quantify how much the addition of the GWs standard siren data can improve the constraints and break the degeneracy on the parametric space of the model, in particular on $b$. For comparison, we show also the constraints for the $\Lambda$CDM model. In particular, we found the following mean values with the relative 68\% CL and 95\% CL errors from the joint analysis (SNe + CC + GWs):
\begin{align}
&H_0 = 69.37^{+0.67(1.45)}_{-0.80(1.41)}\     , \\
&\Omega_{m0} = 0.303^{+0.019(0.038)}_{-0.019(0.037)} \    , \\
&b = 0.383^{+0.134(0.229)}_{-0.116(0.257)}\ .
\end{align}
We note that $b$ tends to be non-null up to the 95\% CL with the addition of GWs data, therefore breaking the degeneracy enough to have a non-null value on $b$ up to 2$\sigma$. This demonstrates the potential of the future standard sirens catalogs in joint analysis with another geometric probes. As a comparison, in \cite{Rafael} it was found $b < 0.50$  and $b<0.13$ at the 95\% CL from CC + $H_0$ and JLA + BAO + CC + $H_0$, respectively, while in \cite{Basilakos} it was found $b < 0.25 $ at the 68\% CL. The data used in these works are from different physical nature and accuracy and, thus, a direct comparison seems not to be appropriate. However, all these constraints are compatible with each other within the 95\% CL. Also, we can note how much the addition of GWs data can improve the constraints on $\Omega_{m0}$ and $H_0$ parameters. 
A detailed analysis from GW standard sirens on other viable $f(R)$ models as well as full discussion will be presented in a future communication. 

\begin{figure}
\begin{center}
\includegraphics[width=3.3in]{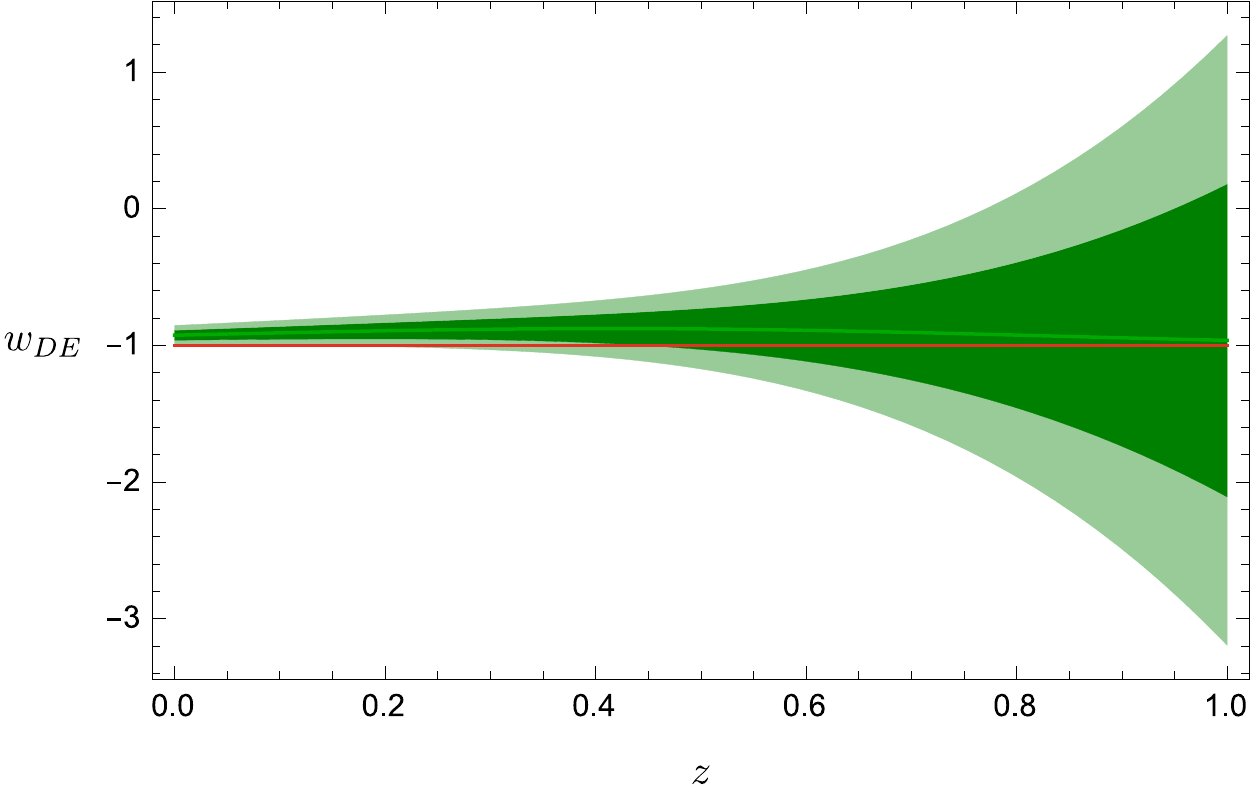}
\caption{Reconstruction of the effective dark energy equation of state parameter at the 68\% and 95\% CL from the SN + CC + GW analysis for the Hu-Sawicki model. The prediction of the $\Lambda$CDM model (red solid line) is shown for comparison.}
\label{fig:wDE}
\end{center}
\end{figure}

Finally, it is interesting to analyze the consequences of our results on the effective dark energy equation of state parameter. This can be expressed as \cite{Basilakos}
\begin{equation}
w_{DE}(a)=\dfrac{-1-\frac{2}{3}a \frac{d\ln{E}}{da}}{1-\Omega_m(a)},
\end{equation}
where $E(a)= H(a)/H_0$ and
\begin{equation}
\Omega_m(a)=\dfrac{\Omega_{m0}a^{-3}}{E^2(a)}.
\end{equation}
In the case of the standard $\Lambda$CDM model, $w_{DE}=-1$ throughout the entire cosmological evolution. In Figure~\ref{fig:wDE}, we show the $1\sigma$ and $2\sigma$ reconstructions of $w_{DE}(z)$ for the Hu-Sawicki model compared to the prediction of the standard $\Lambda$CDM model. The non-vanishing value of $b$ makes the effective dark energy term behave as quintessence at late times for $z < 0.4\ (0.1)$ at the 68\% (95\%) CL. Beyond this range, the effective equation of state is compatible $w_{DE} = -1$ at the 95\% CL. The green line represents the evolution of $w_{DE}$ from the mean values of the MCMC analysis on SNe + CC + GWs.

\section{Final remarks}
\label{Conclusions}

Assuming an effective luminosity distance for GWs, which can be physically interpreted as possible corrections on the GW amplitude propagation between source and detector, we obtained new observational constraints on the running of the Planck mass under appropriate stability conditions (see summary in Table \ref{tab:results}), from 1000 standard siren events from binary neutron star mergers within the ET power spectral density noise. We found that the GW amplitude damping correction can be preferentially non-zero at the 68\% CL and 95\% CL, for  $\alpha_{M0} < 0$, $\alpha_{M0} > 0$, respectively. 

Furthermore, we combined our simulated GW data with the latest available SNe and CC measurements to constrain the parametric Hu-Sawicki gravity model. In doing so, we found that the deviation from GR may be evidenced as non-null at the 95\% CL. These results demonstrate the statistical accuracy that can be achieved by future ground based GW observatory as ET detector. 

The underlying assumption of our analysis is $c_T = c$ at all redshifts. This hypothesis is strongly motivated from the GW170817 constraints to hold locally. On the other hand, the evidence for $c_T = c$ might not be falsified due to the lack the gravitational wave observations at high $z$, even in the near future, since within the LIGO/VIRGO sensitivity, we expect to measure standard siren events from BNS only at very low $z$. 
Thus, as significant deviations from GR are expected only at moderate to high $z$ and/or large scale, it may be interesting to relax the condition $c_T/c = 1$ and perform forecast analyses on the ratio $c_T/c$ from binary systems at high $z$ within the sensitivity of the future detectors such as ET and DECIGO, and check for possible consequences on modified gravity phenomenology.
It would be also interesting to apply a similar methodology as developed here to investigate other well-motivated classes of modified gravity models, as well as to perform joint analysis between standard siren mock events and CMB data (present and future missions). In this way, we will be able to determine the level of deviations from GR as suggested by the combination of future astronomical observations in the next decade. 
\\

\begin{acknowledgments}
\noindent The authors are grateful to Salvatore Capozziello and Jose C. N. de Araujo for useful comments and discussions. R.C.N. would like to thank FAPESP for financial support under the project \# 2018/18036-5. The authors also wish to thank the referee for his/her constructive comments.

\end{acknowledgments}

\begin{widetext}
\section*{Appendix A: PN coefficients}

 For the convenience of the reader, we list below the PN coefficients \cite{TF2} we used for the waveform model Eq.~(\ref{A}) and Eq.~(\ref{phi}).
The individual masses and spin parameters, $m_i$ and $\chi_i$
$(i = 1, 2)$, are encoded in the following parameter combinations:
\begin{align}
&\delta = \ \dfrac{m_1 - m_2}{M}, \\
&\chi_s = \ \dfrac{\chi_1 + \chi_2}{2}, \\
&\chi_a = \ \dfrac{\chi_1 - \chi_2}{2}.
\end{align}
\\
The PN amplitude expansion coefficients are:
\begin{subequations}
\begin{align}
A_0 = &\ 1, \\
A_1 = &\ 0, \\
A_2 = &\ \dfrac{451}{168} - \dfrac{323}{224}, \\
A_3 = &\ \dfrac{27 \delta \chi_a}{8} + \left(\dfrac{27}{8} - \dfrac{11 \eta}{6} \right) \chi_s, \\
A_4 = &\ \dfrac{27312085}{8128512} - \dfrac{1975055 \eta}{338688} + \dfrac{105271 \eta^2}{24192}+ \left( 8\eta + \dfrac{81}{32} \right) \chi^2_a - \dfrac{81}{16} \delta \chi_a \chi_s + \left(\dfrac{81}{32} + \dfrac{17 \eta}{8} \right) \chi^2_s, \\
A_5 = & - \dfrac{85 \pi}{64} + \dfrac{85 \pi \eta}{16} + \delta \left( \dfrac{285197}{16128} - \dfrac{1579 \eta}{4032} \right) \chi_a + \left( \dfrac{285197}{16128} - \dfrac{15317 \eta}{672} - \dfrac{2227 \eta^2}{1008} \right) \chi_s, \\
A_6 = & - \dfrac{177520268561}{8583708672} + \Big( \dfrac{545384828789}{5007163392} - \dfrac{205 \pi^2}{48}\Big) \eta - \dfrac{3248849057 \eta^2}{178827264} 
+ \dfrac{34473079 \eta^3}{6386688} \\
&+ \Big( \dfrac{1614569}{64512} - \dfrac{1873643 \eta}{16128} + \dfrac{2167 \eta^2}{42} \Big) \chi^2_a +
\Big( \dfrac{31 \pi}{12} - \dfrac{7 \pi \eta}{3} \Big) \chi_s + \Big( \dfrac{1614569}{64512} - \dfrac{61391 \eta}{1344} + \dfrac{57451 \eta^2}{4032} \Big) \chi^2_s \\
&+
\delta \chi_a \Big( \dfrac{31 \pi}{12} + \Big( \dfrac{1614569}{32256} - \dfrac{165961 \eta}{2688} \Big) \chi_s \Big) 
\end{align}
\end{subequations}
\noindent The phase $\Phi(f)$ expansion coefficients are:
\begin{subequations}
\begin{align}
\alpha_2 = &\ \dfrac{3715}{756} + \dfrac{55 \eta}{9}, \\
\alpha_3 = & -16\pi + \dfrac{113 \delta \chi_a}{3} + \left( \dfrac{113}{3} - \dfrac{76 \eta}{3} \right) \chi_s, \\
\alpha_4 = &\ \dfrac{15293365}{508032} + \dfrac{27145 \eta}{504} + \dfrac{3085 \eta^2}{72} + \left(200 \eta - \dfrac{405}{8} \right) \chi^2_a - \dfrac{405}{4} \delta \chi_a \chi_s + \left(\dfrac{5 \eta}{2} - \dfrac{405}{8} \right) \chi^2_s, \\
\alpha_5 = &\ [ 1 + \log (\pi M f) ] \left[ \dfrac{38645\pi}{756} - \dfrac{65 \pi \eta}{9} + \delta \left( -\dfrac{140 \eta}{9} - \dfrac{732985}{2268} \right) \chi_a
+ \left( - \dfrac{732985}{2268} + \dfrac{24260 \eta}{81} + \dfrac{340 \eta^2}{9} \right) \chi_s \right], \\
\alpha_6 = &\ \dfrac{11583231236531}{4694215680} - \dfrac{6848 \gamma_E}{21} - \dfrac{640 \pi^2}{3} + \left(\dfrac{15737765635}{3048192} + \dfrac{2255 \pi^2}{12} \right) \eta + \dfrac{76055 \eta^2}{1728} - \dfrac{127825 \eta^3}{1296} \\
&- \dfrac{6848}{63} \log(64 \pi M f) + \dfrac{2270}{3} \pi \delta \chi_a \left( \dfrac{2270 \pi}{3} - 520 \pi \eta \right) \chi_s, \\
\alpha_7 = &\ \dfrac{77096675 \pi}{254016} + \dfrac{378515 \pi \eta}{1512} - \dfrac{74045 \pi \eta^2}{756} + \delta \left(- \dfrac{25150083775}{3048192} + \dfrac{26804935 \eta}{6048} - \dfrac{1985 \eta^2}{48} \right)\chi_a \\
&+ \left( -\dfrac{25150083775}{3048192} + \dfrac{10566655595 \eta}{762048} - \dfrac{1042165 \eta^2}{3024} + \dfrac{5345 \eta^3}{36} \right) \chi_s.
\end{align}
\end{subequations}

\section*{Appendix B: SNe and CC data set}
We provide here some details of the low-redshift cosmological observables that we used to complement the GW mock data in the statistical analysis on the HS model. 
\\

The first data set is the Pantheon sample \cite{Scolnic18}, composed of 1048 Supernovae (SNe) Ia in the redshift range $0.01<z<2.3$. In this compilation, all the SNe are standardized through the SALT2 light-curve fitter, in which the distance modulus is modelled as follows \cite{SALT2}:
\begin{equation}
\mu=m_B-M+\alpha x_1-\beta C + \Delta_M +\Delta_B ,
\label{mu_SALT2}
\end{equation}
where $m_b$ is the $B$-band apparent magnitude of each SN and $M$ is its
absolute magnitude, while $\Delta_M$ and $\Delta_B$ account for the host-mass galaxy and the distance bias corrections, respectively. Moreover, $x_1$ and $C$ are the stretch and color parameters of each SN light-curve, respectively, with their relative coefficients $\alpha$ and $\beta$. On the other hand, the distance modulus predicted by a cosmological model is given as
\begin{equation}
\mu(z)=5\log_{10}\left[\dfrac{d_L(z)}{1\text{ Mpc}}\right]+25.
\end{equation}
As shown in \cite{Riess18}, under the assumption of a flat universe one can compress the full SN sample into a set of 6 cosmological model-independent measurements of $E(z)^{-1}$, where $E(z)\equiv H(z)/H_0$ is the dimensionless Hubble parameter. This approach allows us to properly marginalize over the SN nuisance parameters in the fit. Thus, taking into account the correlations among the $E^{-1}(z)$ measurements, we can write the likelihood function associated to the SN data as \cite{DAgostino19}
\begin{equation}
\mathcal{L}_\text{SN}\propto \exp\left[-\dfrac{1}{2}\mathbf{A}^\text{T} \mathbf{C}_\text{SN}^{-1} \mathbf{A}\right],
\end{equation}
where $\mathbf{A}\equiv E^{-1}_{obs,i}-E^{-1}_{th}(z_i)$ quantifies the difference between the measured values and the values predicted by a cosmological model, and $\mathbf{C}_\text{SN}$ is the covariance matrix resulting from the correlation matrix given in \cite{Riess18}. 
\\

The second data set is built upon the differential age approach \cite{Jimenez02}, which represents a model-independent method to characterize the expansion of the universe up to $z<2$.  In this technique, passively evolving red galaxies are used as cosmic chronometers (CC) to measure the age difference $(dt)$ of the universe at two close redshifts $(dz)$. Thus, one can estimate the Hubble parameter as
\begin{equation}
H(z)=-\dfrac{1}{(1+z)}\dfrac{dz}{dt} .
\end{equation}
In our analysis, we used the compilation of 31 $H(z)$ uncorrelated measurements collected in \cite{Capozziello18} (see references therein). We can then write the likelihood function relative to the CC data as
\begin{equation}
\mathcal{L}_\text{CC}\propto\exp\left[-\dfrac{1}{2}\displaystyle{\sum_{i=1}^{31}}\left(\dfrac{H_{obs,i}-H_{th}(z_i)}{\sigma_{H,i}}\right)^2\right] ,
\end{equation}
where $H_{obs,i}$ are the observed measurements with their relative uncertainties $\sigma_{H,i}$, while $H_{th}(z_i)$ are the theoretical values of the Hubble parameter estimated from using a specific cosmological model.

\end{widetext}

\end{document}